\documentclass[
    reprint,
    amsmath,amssymb,
    aps,
    unsrtbib,
    nofootinbib,
    floatfix,
]{revtex4-2}

\usepackage{placeins}
\usepackage{hyperref}
\hypersetup{
pdfborder={0 0 1},
	colorlinks=true,
	linkcolor=blue,
	citecolor=blue,
	urlcolor=blue,
}
\usepackage{graphicx}
\usepackage{dcolumn}
\usepackage{bm}

\usepackage{tikz}
\usetikzlibrary{positioning,arrows.meta}

\definecolor{lightblue}{rgb}{0.1, 0.5, 1.0}

\begin{document}

\title{Hybrid Neural Simulation-Based Inference for Robust Applications and Limited-Budget Scenarios}

\author{Sean Benevedes}
\email{sbenevedes3@gatech.edu}
\affiliation{School of Physics, Georgia Institute of Technology, Atlanta, GA 30332, USA}

\author{Mani Dehghan}
\email{mdehghan3@gatech.edu}
\affiliation{School of Physics, Georgia Institute of Technology, Atlanta, GA 30332, USA}

\author{Aishik Ghosh}
\email{AishikGhosh@physics.gatech.edu}
\affiliation{School of Physics, Georgia Institute of Technology, Atlanta, GA 30332, USA}
\affiliation{Lawrence Berkeley National Laboratory, Berkeley, CA 94720, USA}

\author{Tae Hyoun Park}
\email{taehyoun@mpp.mpg.de}
\affiliation{Max Planck Institute for Physics, Boltzmannstr.~8, 85748 Garching, Germany}

\begin{abstract}
We develop two hybrid techniques that approach the performance of neural simulation-based inference (NSBI) analyses while substantially reducing the computational cost of inference and preserving some or all of the reliability guarantees of parametric methods. The first approach is broadly applicable, while the second is tailored to a class of particle physics analyses that admit a semi-parametric NSBI formulation.

With only a modest compromise in raw sensitivity, these methods represent an important step toward computationally efficient NSBI in offline analyses and also open the door to the exploration of trigger-level applications in the future. Based on our comparison studies, we recommend the use of our first approach, Latent Categories, for robust and efficient inference. 
\end{abstract}
\maketitle
\section{Introduction}
\label{sec:intro}

Modern particle physics experiments record increasingly complex, high-dimensional detector data. A single collision event may contain information from thousands to millions of detector channels, from which physicists must extract subtle statistical signatures to test hypotheses about fundamental interactions. The central challenge is therefore to construct statistical procedures that retain as much information as possible from these rich observations while remaining computationally tractable and sufficiently reliable for precision measurements.

Traditionally, particle physics analyses compress detector information into one or a small number of high-level observables, whose distributions are compared with predictions using binned likelihood fits. While this approach is computationally efficient and well understood, such low-dimensional summaries are generally lossy. Recent advances in machine learning have instead enabled analyses that operate directly on unbinned, high-dimensional data by learning probability densities or density ratios from simulated events. These learned quantities can be used to construct likelihoods and test statistics within standard frequentist inference frameworks, substantially reducing the information loss associated with handcrafted summary observables.

These methods, collectively referred to as neural simulation-based inference (NSBI)~\cite{Cranmer:2015bka,doi:10.1073/pnas.1912789117}, have demonstrated improved statistical sensitivity in a variety of settings, well beyond particle physics~\cite{Brandes:2024vhw,Dax:2021tsq,SimBIG:2023ywd}. They are particularly advantageous when no low-dimensional parameter-independent sufficient statistic exists, for example in measurements sensitive to non-linear quantum interference effects~\cite{Ghosh_jrjc_proc} or in multi-parameter measurements such as Effective Field Theory (EFT) analyses~\cite{Brehmer:2018eca}. Until recently, however, deploying these techniques in experimental particle physics while maintaining the precision and robustness required for real analyses remained a significant challenge. Unlike in parametric methods, the estimated density ratios from networks may be poorly calibrated or biased. 

A major step toward practical NSBI was the semi-parametric framework introduced by the ATLAS Collaboration~\cite{ATLAS:2025clx}, which was subsequently used to obtain a substantially improved measurement of the Higgs boson width compared to a conventional histogram-based analysis~\cite{ATLAS:2024jry}. In this framework, neural networks estimate event-level density ratios that are combined analytically to construct likelihoods across the full space of parameters of interest and nuisance parameters. This formulation enables density-ratio estimators to be trained and rigorously validated at the precision required for experimental analyses while naturally incorporating the propagation of systematic uncertainties. Ref.~\cite{Ghosh:2025fma} further extends this semi-parametric approach to incorporate the use of matrix-element information in training, based on core concepts from Refs.~\cite{Brehmer:2018eca,Brehmer:2019xox}.

The computational cost of this approach, however, remains substantial during both training and, crucially, inference~\cite{refId0}. A realistic analysis may require training and evaluating thousands of neural networks to achieve the precision required for reliable inference. A recent proposal~\cite{Benevedes:2025nzr} to use parametrized ensembles to quantify neural network uncertainties could reduce these requirements, but it still requires ensembling multiple models and a nontrivial well-specification assumption. Evaluating these models for every event across many hypotheses for both the parameters of interest and nuisance parameters demands substantial memory and computational resources. These requirements already pose a challenge for many offline analyses and currently preclude deployment in resource-constrained environments such as software trigger systems. High-dimensional analysis techniques deployed at the trigger, however,
would unlock rate-limited analyses with subtle signal signatures, such as those arising from interference effects, for which it is difficult to define conventional triggers.

This motivates the development of hybrid inference strategies that preserve much of the statistical power of NSBI while substantially reducing the computational cost of inference. One approach in this direction was taken in Ref.~\cite{Brehmer:2018eca}, which introduced a locally optimal observable that recovers the sensitivity of NSBI in the vicinity of the Standard Model (SM). However, this local approximation does not generally maintain optimal performance throughout the region of parameter space relevant for the analysis.

In this work, we introduce two hybrid machine learning approaches that combine the computational efficiency and various reliability guarantees of parametric inference with the information-preserving capabilities of high-dimensional neural networks, without relying on local approximations. The first method, which we will call Latent Categories (LC), is applicable to a broad class of NSBI problems. In LC, a network learns to automatically partition events into categories in its latent space using a learning objective that maximizes the information between this categorization and a parameter of interest. Inference can then be performed exactly as in a conventional histogram-based parametric analysis, but this categorization generically does not correspond to the binning of any single observable, allowing it to approach the power of NSBI even where no parameter-independent low-dimensional sufficient statistic exists. 

The second method, called Mixture of Summary Statistics (MSS), is designed for a class of particle physics analyses admitting a semi-parametric NSBI formulation. It approximates process-level contributions to event likelihood ratios using histograms, allowing the full likelihood ratio to be approximated analytically while dramatically reducing inference cost. We find that the first approach sacrifices very little sensitivity in exchange for broad applicability and robust likelihood estimation, whereas the second achieves similar sensitivity on its restricted class of problems, but with fewer robustness guarantees, relying on well-trained (but not necessarily well-calibrated) classifiers.

These techniques provide a practical path toward computationally efficient high-dimensional inference in particle physics. The LC approach in particular significantly reduces the computational burden of NSBI while retaining most of its statistical advantages, making sophisticated machine-learning-based inference feasible across offline analyses, and opening the door to exploration of resource-constrained online applications such as in software triggers and data scouting / trigger-level analyses at the LHC~\cite{LHCb:2018zdd,CMS:2024zhe}.

The remainder of this paper is organized as follows. Section~\ref{sec:methods} introduces the baseline NSBI implementation and the two proposed hybrid inference methods, and Section~\ref{sec:datasets} describes the datasets studied. Section~\ref{sec:results} evaluates statistical performance on both a Gaussian benchmark and a Higgs physics analysis. Finally, Section~\ref{sec:discussion} discusses the strengths and limitations of each method, the broader implications of computationally efficient hybrid inference, and concludes the paper.

\section{Methods}
\label{sec:methods}

We compare four inference strategies: a traditional histogram-based analysis, an NSBI approach, and the two proposed novel hybrid methods.

\subsection{Histogram}
\label{sec:histogram}

The histogram analysis serves as the conventional baseline. Events are summarized by a locally optimal or physically motivated one-dimensional observable. Simulated events are used to construct binned templates for each value of the parameter of interest, and inference is performed using a binned Poisson likelihood over the histogram bins.

\subsection{NSBI}
\label{sec:NSBI}

Neural network classifiers trained with a binary cross-entropy loss can be used to estimate probability density ratios~\cite{Cranmer:2015bka}. By conditioning the classifier on the parameter of interest $\theta$, they can be extended to estimate conditional density ratios of the form $\frac{p(x_i|\theta)}{p(x_i|\mathrm{ref})}$, where $\mathrm{ref}$ denotes a fixed reference hypothesis. The semi-parametric NSBI implementation introduced in Ref.~\cite{ATLAS:2025clx} further decomposes this density ratio as

\begin{equation}
\label{eq:so_mm}
\frac{p(x_i|\mu)}{p(x_i|\mathrm{ref})}
=
\frac{1}{\nu(\mu)}
\sum_{J=1}^{C_{\mathrm{proc}}}
f_J(\mu)\,\nu_J\,
\frac{p(x_i|J)}{p(x_i|\mathrm{ref})}.
\end{equation}

Equation~(\ref{eq:so_mm}) expresses the conditional density ratio as a weighted sum of process-level density ratios, enabling each component to be estimated independently and recombined analytically for arbitrary values of $\mu$.

Here, $x_i$ denotes the vector of observables describing an event, $\mu$ is the parameter of interest in this physics context, $\nu_J$ is the Standard Model expected event yield for process $J$, $\nu(\mu)$ is the total expected event yield for a given value of $\mu$, $f_J(\mu)$ describes the dependence of each process on $\mu$ according to the underlying theory model, and $C_{\mathrm{proc}}$ is the total number of contributing physics processes.

For the Gaussian example, a parametrized NSBI implementation is sufficient. For the Higgs physics example, we instead use the semi-parametric formulation to maximize sensitivity, following the ATLAS implementation. Simulated samples from the off-shell Higgs boson analysis in the four-lepton final state from Ref.~\cite{Ghosh:2025fma} are used for a signal strength measurement similar to Ref.~\cite{ATLAS:2025clx}, resulting in three physics processes with distinct scaling functions $f_J(\mu)$.

\begin{figure*}[t]
    \centering
    \includegraphics[width=\textwidth]
    {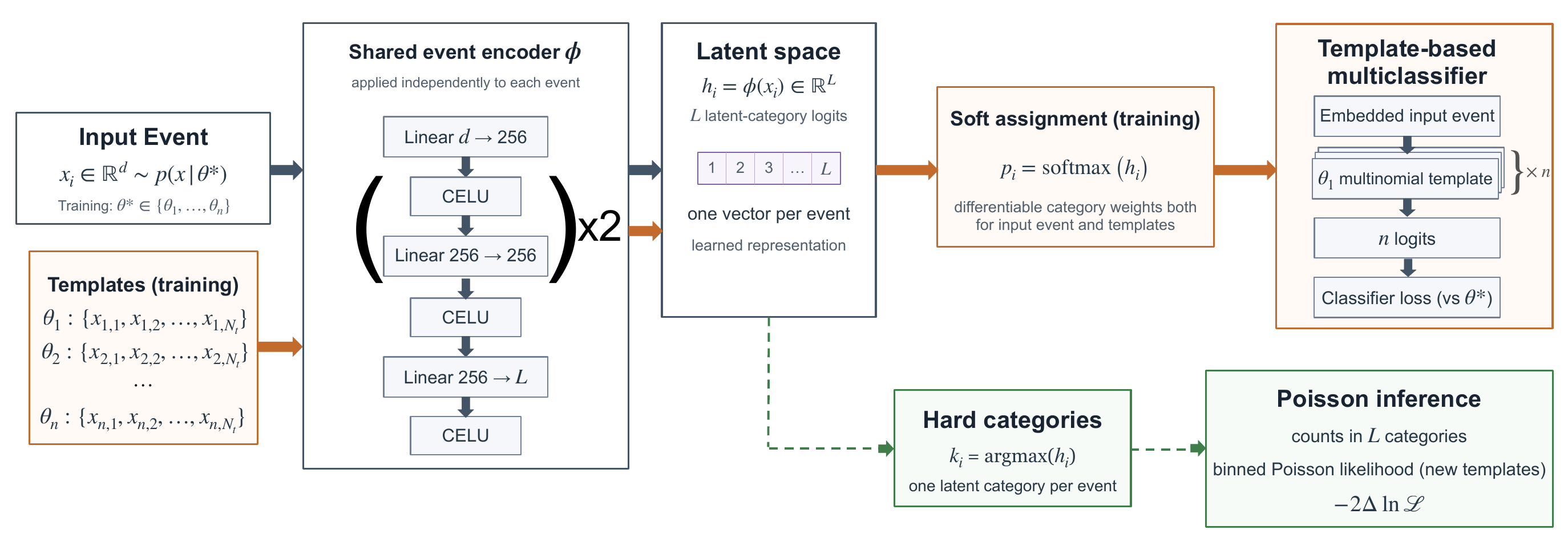}
    \caption{Schematic diagram of the training and inference pipelines for the Latent Categories approach. Black indicates components of the architecture present at both training and inference time. Orange indicates training-specific elements, showing how the encoder is trained as a multiclassifier, and green shows the inference pipeline.}
    \label{fig:SchematicLatentCat}
\end{figure*}

\subsection{Latent Categories}
\label{sec:LatestCategories}

The Latent Categories (LC) approach learns a latent categorical representation of high-dimensional events that preserves the information required for likelihood-ratio estimation. Rather than performing inference directly with a neural classifier, the learned latent representation defines event categories on which a conventional Poisson likelihood is constructed. A schematic diagram of the procedure is shown in Fig.~\ref{fig:SchematicLatentCat}.

The model takes an event as input and (after a softmax) yields an $L$-dimensional, normalized, positive output, where $L$ is the number of latent categories to be learned. The training dataset for the model consists of $N_\text{train}$ events, equally distributed among $n$ classes. Each class corresponds to a value of the parameter $\theta$, and the events in each class are drawn at the corresponding value of $\theta$. The model is then trained as a supervised multiclassifier over events using the categorical cross-entropy loss, where the labels are the value of $\theta$ at which each event was drawn. In our experiments, we will take $n=21$, with the corresponding parameters being uniformly spaced between $\theta_\text{min}$ and $\theta_\text{max}$ for each experiment. This corresponds to a grid spacing of $0.1$ for the Gaussian case study and $0.2$ for the physics case study.

The key step is to design a classifier head that takes the $L$-dimensional latent representation of an event generated at one of the $n$ parameter values and outputs the probability that it was generated at each of them. This classifier head should be inference-aware, in the sense that it should promote learning a latent representation that is directly useful for the downstream binned analysis. To accomplish this, at each training step we first construct differentiable category templates for every parameter point. A random batch of simulated training events is passed through the encoder, and the resulting soft category assignments are summed with the appropriate event weights to estimate the expected occupancy of each latent category.

Separately, we generate a batch of events, each drawn at a randomly selected parameter value. The events are passed through the same encoder, yielding a differentiable soft count vector for each event. For each event, we then evaluate its multinomial log-likelihood at each of the $n$ parameter points using the corresponding latent category templates. These $n$ log-likelihood values serve directly as the logits of the parameter classifier.

We note for completeness that another inference-aware procedure would be one in which the network is trained to classify pseudo-experiments of the desired size, rather than individual events. Since this training loop would proceed over aggregated pseudo-experiments rather than individual events, it could be thought of as training a DeepSet model~\cite{DBLP:journals/corr/ZaheerKRPSS17} with the embedding model as $\phi$, summation as the aggregator, and the log-likelihood computation using the Poisson templates as $\rho$. This approach could then be compared to previous dataset-level NSBI approaches, such as those of Refs.~\cite{Heinrich:2023bmt,wehenkel2026justtakestwoscaling}. In our experiments, we find that the per-event classification procedure performs at least as well as this pseudo-experiment procedure, so we present results for the simpler per-event prescription.

During inference, the softmax layer used to produce the $L$-dimensional embedding is replaced by an argmax\footnote{One could also imagine other ways of assigning each event to one category based on soft scores; in particular, our training procedure more naturally corresponds to sampling the category with probability equal to the softmax outputs. However, we find slightly better performance with argmax in our experiments, and argmax carries the benefit of yielding a deterministic analysis.} operation so that each event is assigned uniquely to a single latent category. This means that we have effectively learned a binning, though not one generically corresponding to a histogram of any one-dimensional parameter-independent observable; indeed, the categories do not even admit a natural notion of ordering. Inference then proceeds exactly as in a histogram-based analysis, using independent Poisson likelihoods for each category.

The training procedure is then inference-aware in the sense that the template-likelihood approach to assigning classification scores mirrors the downstream inference procedure. More precisely, in the limit of exact simulation-derived templates and hard category assignments, the training objective selects the $L$-category binning that maximizes the expected log-posterior of the true generating parameter given a single event's category, and therefore the mutual information between the parameter of interest and that category. For a discretized parameter space this statement is exact, and in the limit of fine parameter spacing this discrete result approaches the corresponding continuum result. This categorization approach is inspired by inference-aware work in Ref.~{\cite{DeCastro:2018psv}}. In contrast with prior work on learned event binnings~\cite{Simpson:2022suz,Wunsch:2020iuh}, LC searches for a binning which is maximally informative for parameter inference not only locally, in the vicinity of a reference parameter value, but globally over the training parameter space.

\subsection{Mixture of Summary Statistics}
\label{sec:Histogram_MM}

\begin{figure*}[t]
    \centering
    \includegraphics[width=\textwidth]
    {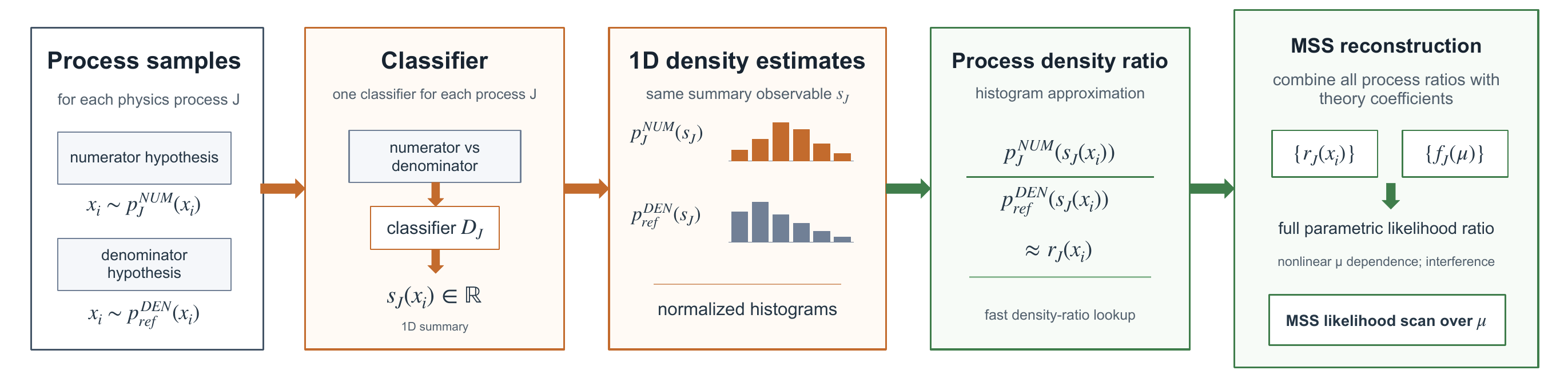}
    \caption{Schematic diagram of the training and inference pipelines for the Mixture of Summary Statistics approach. As with Fig.~\ref{fig:SchematicLatentCat}, orange shows how the model is trained and green shows how it is used for inference.}
    \label{fig:SchematicMSS}
\end{figure*}

The Mixture of Summary Statistics (MSS) approach exploits the factorization in Eq.~(\ref{eq:so_mm}) to replace neural density-ratio estimators with histogram-based density estimates while retaining the semi-parametric likelihood formulation. A schematic diagram of this procedure is shown in Fig.~\ref{fig:SchematicMSS}. For each physics process, a dedicated classifier is trained to distinguish events generated under the numerator and denominator hypotheses. The classifier output serves as a one-dimensional summary statistic from which normalized histograms are constructed for both hypotheses. Evaluating the histogram densities at the classifier response provides an estimate of the corresponding process-level density ratio.

\begin{equation}
\label{eq:hmm_full}
\frac{p(x_i|\mu)}{p(x_i|\mathrm{ref})}
\approx
\frac{1}{\nu(\mu)}
\sum_{J=1}^{C_{\mathrm{proc}}}
f_J(\mu)\,\nu_J\,
\frac{\hat{p}_J(s_J(x_i))}{\hat{p}_{\mathrm{ref}}(s_J(x_i))}.
\end{equation}

This approximation relies on the fact that a well-trained classifier is known to provide a sufficient summary statistic to estimate density ratios~\cite{Cranmer:2015bka}, and that both the numerator and denominator densities are expressed in terms of that summary observable. However, unlike the NSBI approach in Sec.~\ref{sec:NSBI}, the classifier here does not need to be well calibrated (a requirement that is typically more challenging to achieve and necessitates the use of ensembles of networks). This yields a fully factorized approximation of the NSBI likelihood in which all learned components are replaced by one-dimensional density estimators. In practice, we find that it is important to validate that the networks have summarized sufficiently well, for instance with re-weight diagnostics~\cite{ATLAS:2025clx}, even though calibration is not required and normalization is guaranteed. Coarse binning may also lead to an insufficient summarization and loss of robustness.

Unlike a conventional histogram analysis, these histograms are used as density estimators rather than Poisson templates and provide per-event likelihood ratios. The resulting process-level density ratios are combined using the theory-dependent weights $f_J(\mu)$, preserving the full parameter dependence of the semi-parametric NSBI formulation, including non-linear effects such as quantum interference.

\section{Datasets}
\label{sec:datasets}

We demonstrate the techniques on a Gaussian example and a physics dataset. 

A five-dimensional Gaussian dataset is generated with a mean at the origin and a covariance matrix

\begin{equation}
    \begin{pmatrix}
        1 & a \theta & 0 & 0 & 0 \\
        a \theta & 1 & 0 & 0 & 0 \\
        0 & 0 & 1 & b \theta & 0 \\
        0 & 0 & b \theta & 1 & 0 \\
        0 & 0 & 0 & 0 & 1
    \end{pmatrix}
\end{equation}

where $\theta$ is the parameter of interest that will be measured and $a, b$ are parameters. Samples are generated for $\theta$ in the range $[-1,1]$, so we must have $|a|,|b|<1$ to maintain a positive-definite covariance matrix. We choose this example because for generic values of $a$ and $b$, despite its simplicity, it does not admit a one-dimensional $\theta$-independent summary statistic sufficient for optimal inference on $\theta$ over its entire range. This can be seen from the fact that the log-likelihood\footnote{Or more precisely, the terms in the log-likelihood which are not constant with respect to $\theta$.} depends on the combinations $x_0^2 + x_1^2$, $x_0 x_1$, $x_2^2 + x_3^2$, and $x_2 x_3$, and the relative weights of these combinations are $\theta$ dependent.

We choose $a = 0.8$ and $b = 0.4$ as our benchmark point, and we consider inference on $\theta$ with experiments of (mean) size $50$. We generate $300,000$ events at each of $11$ parameter values equally spaced between $\theta =-1$ and $\theta=1$ for a total of $3.3$ million events, retaining $70\%$ of the data for training, $15\%$ for validation, and $15\%$ for inference tests. As with the below physics case study, we obtain samples for arbitrary values of the parameter via event reweighting.

The physics dataset is taken from Ref.~\cite{Ghosh:2025fma} and represents off-shell Higgs production and decay to a four-lepton final state, including interference effects with the background, $gg (\rightarrow H^*) \rightarrow ZZ \rightarrow 4l$. It is used to study a signal strength measurement, following the ATLAS NSBI methodology paper~\cite{ATLAS:2025clx}. The simulations include events generated according to the signal-only (S), background-only (B), interference-only (I), and full signal-background-interference (SBI) contributions, with several million weighted events generated for each scenario at the Standard Model (SM) signal strength, $\mu=1$. Each event contains the full matrix-element information needed to reweight between these scenarios and to arbitrary values of $\mu$; we consider $\mu \in [0,4]$. The simulated samples are split into training, validation, and test datasets (with $70\%, 15\%,$ and $15\%$ of the data, respectively) before reweighting to prevent data leakage, and at $\mu=1$ the expected number of events in each inference experiment is $564$ (but this expected yield depends on $\mu$, unlike the Gaussian example). 

\section{Results}
\label{sec:results}
We study the relevant techniques for a Gaussian and a physics example dataset.

\subsection{Gaussian example}
\label{sec:GaussianExample}

For the Gaussian example, we compare LC against NSBI, implemented using a conditioned neural network density-ratio estimator, and a traditional histogram approach.
In order to compare to the strongest possible one-dimensional traditional histogram approach, we choose the binned feature to be $a x_0 x_1 + b x_2 x_3$, which is the \textit{score}, the derivative of the log-likelihood with respect to the parameter, evaluated at $\theta = 0$ and therefore a locally sufficient statistic about that parameter value~\cite{Atwood:1991ka,Davier:1992nw,Diehl:1993br,Nachtmann:2004fy,Brehmer:2019xox}.
Since we are primarily concerned with power, i.e. with achieving the smallest valid confidence intervals possible, our figure of merit will be the expected $-2\Delta \ln \mathcal L$ evaluated with the true value of $\theta$ assumed to be $0$.
The score feature that we choose is then the locally optimal $\theta$-independent statistic for $\theta$ inference, though we would expect a priori that its performance will suffer away from $\theta = 0$, and that its $-2 \Delta \ln \mathcal L$ will be below those of the other approaches in that regime.
We use $32$ bins, which we find to be sufficiently fine-grained to approach maximal sensitivity of the binned analysis.

Our LC implementation uses three hidden layers with width $256$, CELU activations~\cite{DBLP:journals/corr/Barron17a}, and $L=48$ latent categories. It is trained using the Adam optimizer~\cite{kingma2017adammethodstochasticoptimization} with learning rate $2\times10^{-3}$ for $100$ epochs each consisting of $170$ steps, and each step classifying $12,800$ labeled events using templates of $30,000$ events. Every $5$ epochs, the mean value of $-2\Delta \ln \mathcal{L}$ as evaluated at $\theta \in \{-1.0, -0.5, 0.5, 1.0\}$ is computed on held-out validation data as a figure of merit, and the checkpoint maximizing this metric is selected as the nominal model. To ensure that we present results corresponding to a typical training, we train the network three times with distinct random seeds, and we present results for the median-performing network. We find that this spread from training-to-training is negligible.

Our NSBI implementation uses an ensemble average of $10$ trained networks, each also consisting of three hidden layers with width $256$ and CELU activations. The reference is taken to be $\theta = 0$, so training samples $(x, \theta)$ are taken at each of the other $n-1 = 20$ grid points, and training proceeds as binary classification using binary cross-entropy. The networks are each trained for $80$ epochs on resamplings of the training data, with the best epoch by binary cross-entropy on the validation set selected for each network.

\begin{figure}[htbp]
    \centering
    \includegraphics[width=\linewidth]
    {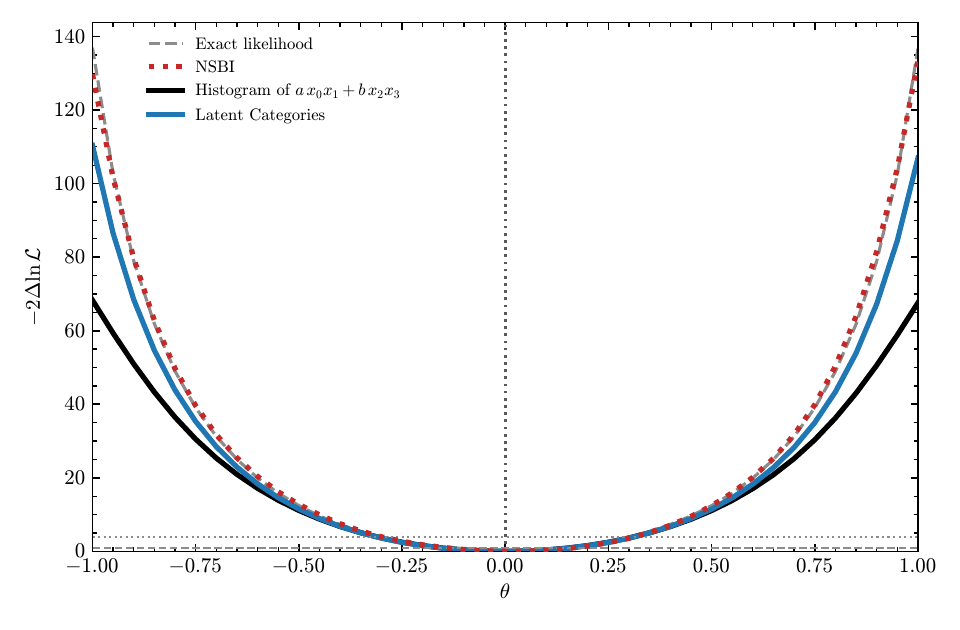}
    \caption{Sensitivity of LC, NSBI, and histogram analyses on the Gaussian example dataset. Narrower curves indicate better sensitivity. The true value of the parameter is shown with a vertical dotted line, and the $1\sigma$ and $2 \sigma$ thresholds are shown with the horizontal dashed and dotted lines.}
    \label{fig:toy_sensitivity}
\end{figure}

Results of the comparison are shown in Fig.~\ref{fig:toy_sensitivity}. As is to be expected, NSBI provides the maximum sensitivity, indicated by the narrowest curves, and it is visually indistinguishable from the line corresponding to using the exact likelihood. However, it is not protected from mismodeling of the likelihood, which we have mitigated in this simple case study through the use of ensembling.  LC is not quite as sensitive as NSBI and the exact likelihood, but it can be seen that its curve is close to that of the exact likelihood throughout the parameter space, and that away from $0$ it indeed outperforms the binned optimal observable.

As such, in this case study, LC successfully improves upon the baseline of the optimal histogrammed analysis while retaining the robustness guarantees of a binned analysis. Moreover, since network training only affects the power and not the reliability of LC inferences, fewer training resources can be used: in this case study, the LC networks take an order of magnitude less compute to train than their NSBI ensemble counterparts. Details about the learned latent categorization can be found in Appendix~\ref{sec:AppendixLatentCategory}.

\subsection{Off-shell Higgs example}
\label{sec:PhysicsExample}

For the physics dataset, we compare the semi-parametric NSBI approach designed by ATLAS with our LC, MSS, and a histogram approach. The histogram is constructed using $m_{4l}$, the mass of the four-lepton system, with 32 bins. 

As with the Gaussian case study, we train LC using three hidden layers with width $256$ and $L=48$ latent categories. We train the model with the Adam optimizer with learning rate $3 \times 10^{-4}$ for $300$ epochs of $150$ steps each. Each step consists of classification of $55,680$ events with templates constructed from $60,000$ events. Every $5$ epochs, the mean value of $-2 \Delta \ln \mathcal{L}$ is evaluated at $\mu \in \{0, 2, 3, 4\}$ on held-out validation data, and the checkpoint maximizing this metric is retained. We again train three networks with distinct random seeds and present results for the median-performing network, again finding that training-to-training spread is negligible.

Our NSBI and MSS implementations use the same networks. Unlike the Gaussian case study, rather than ensembling relatively weak classifiers, we find that it works well to instead train individual strong classifiers. We train one such classifier to learn the signal-to-background ratio and another to learn the SBI-to-background ratio, which together suffice to reconstruct the parametrized likelihood provided in Eq.~\eqref{eq:so_mm} and Eq.~\eqref{eq:hmm_full}. We take each of these networks to have $17$ hidden layers with a width of $1024$ and SiLU activations~\cite{DBLP:journals/corr/HendrycksG16,DBLP:journals/corr/ElfwingUD17,DBLP:journals/corr/abs-1710-05941}. 

We train them with the NAdam optimizer~\cite{dozat.2016} and a learning rate of $10^{-4}$ for $500$ epochs. The learning rate is reduced by a factor of $10$ after five epochs without a reduction in the validation loss, and the checkpoint with the lowest validation loss is retained. Early stopping terminates the training after $20$ epochs without a reduction in the validation loss, but this does not occur in any of our training runs. In a fully realistic analysis, one would likely still ensemble these classifiers to ensure robustness~\cite{ATLAS:2025clx}, but in this fully controlled, synthetic setting we find that one well-trained network suffices. As discussed in App.~\ref{sec:AppendixCoverage}, we find that the coverage performance of MSS is sensitive to the choice of binning, with fine-grained binnings performing better than coarse-grained binnings, so we use $200$ bins.

\begin{figure}[htbp]
    \centering
    \includegraphics[width=0.5\textwidth]
    {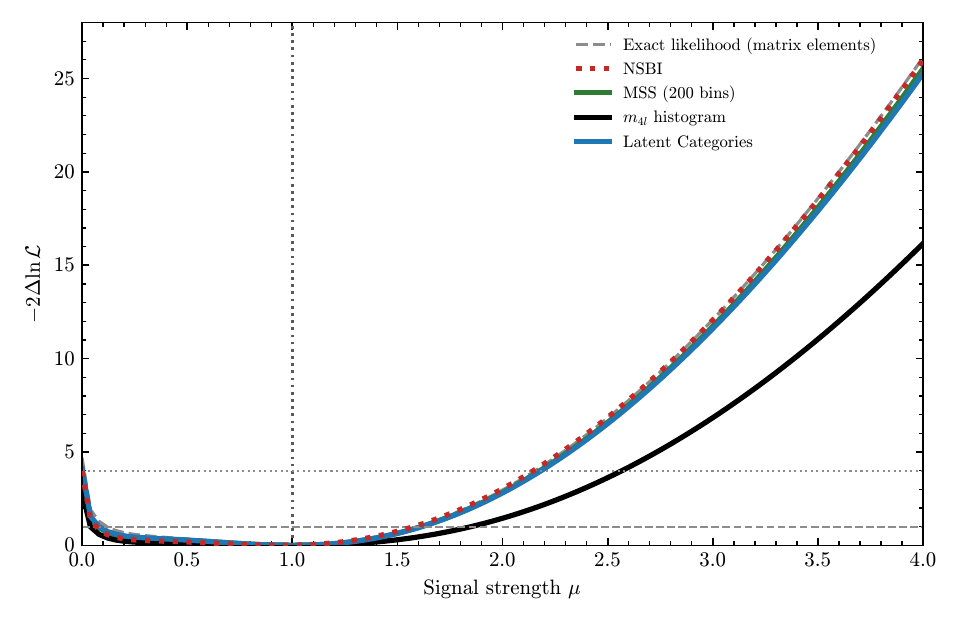}
    \caption{Sensitivity of LC, MSS, NSBI, and histogram analyses on the off-shell Higgs dataset. Narrower curves indicate better sensitivity. The true value of the parameter is shown with a vertical dotted line, and the $1\sigma$ and $2 \sigma$ thresholds are shown with the horizontal dashed and dotted lines.}
    \label{fig:PhysicsResult}
\end{figure}

The results of the comparison are shown in Fig.~\ref{fig:PhysicsResult}. NSBI again nearly saturates the power of the optimal analysis performed with the exact likelihood (which we can access here due to our knowledge of the matrix elements). In this case, both MSS and LC yield almost identical power to NSBI, and the binned $m_{4l}$ analysis is substantially less performant, as expected.

In this case study, MSS and LC both improve substantially upon the binned baseline while retaining improved robustness properties relative to NSBI. LC in particular has the same coverage properties as a conventional binned analysis, and in this case its network takes two orders of magnitude less compute to train than the classifiers used to perform NSBI and MSS. Once again, details about the learned latent categorization can be found in Appendix~\ref{sec:AppendixLatentCategory}.

\section{Discussion and Conclusion}
\label{sec:discussion}

We study two hybrid techniques that approach sensitivities comparable to neural simulation-based inference, without needing to rely on well-calibrated neural estimation of likelihoods or likelihood ratios. A reliable application of NSBI currently requires training thousands of networks and performing inference on supercomputers with terabytes of memory~\cite{ATLAS:2025clx,refId0}. The techniques developed in this work significantly reduce computational cost and can in some cases be run on personal computers, with only a minimal sacrifice in sensitivity. One of them also has all the robustness guarantees of a traditional histogram analysis and is therefore our recommended approach for robust NSBI applications and limited-budget scenarios.

In this approach, a network learns to sort events into a set of learned latent categories, which subsequently enables robust inference using analytical functions. This approach shows promise on both the Gaussian example and physics dataset. Diagnostic checks indicate that more advanced training techniques could further improve performance, both by training more powerful embedding networks through brute force and by better aligning the soft-count training loop with the hard-count by annealing the softmax temperature or using tricks like the straight-through estimator~\cite{DBLP:journals/corr/BengioLC13}. Although such training techniques are computationally more expensive, they could nevertheless reduce the computational cost of inference and the human effort required to validate and calibrate non-parametric NSBI methods. 

We report a two-order-of-magnitude improvement in training efficiency using LC, and we expect the gains over semi-parametric NSBI to be significantly larger in realistic settings with numerous physics processes and systematic uncertainties, where thousands of networks are currently required. An even greater bottleneck for unbinned NSBI analyses is statistical inference, which requires tracking every event under every systematic variation. Our approach substantially reduces the associated computational cost and memory footprint, bringing them in line with those of a traditional histogram-based analysis. In the presence of systematic uncertainties, the latent categorization can be learned from simulated samples generated over a grid of nuisance-parameter values for the few most impactful sources of uncertainty~\cite{ghosh_uncertainty_2021,Louppe:2016ylz}. LC could also be combined with the method of Ref.~\cite{Alvarez:2026umb} to capture systematic uncertainties that do not correspond to known nuisance parameters but can be estimated by comparing multiple simulations.

The second approach we study introduces a mixture of summary statistics that exploits the semi-parametric factorization underlying the ATLAS NSBI framework~\cite{ATLAS:2025clx}, replacing neural density-ratio estimators with per-process histogram density estimates. On the off-shell Higgs dataset, this approach also nearly reproduces the sensitivity of full NSBI while removing the need for well-calibrated neural networks at inference time. It still requires well-trained networks and sufficiently fine histogram binning for robustness. Therefore, NSBI diagnostic checks and coverage tests are still required to validate that the summarization of the high-dimensional data is sufficient. 

While this technique could become a useful alternative approach in environments with strict computational budgets, assessing its robustness for real-world problems would require access to more realistic NSBI benchmark datasets, such as the simulated samples used to develop the ATLAS methodology~\cite{ATLAS:2025clx}. The release of such datasets would therefore enable the wider community to develop improved neural inference techniques better aligned with the needs of particle and nuclear physics experiments. The applicability of this approach is currently limited to analyses that admit a semi-parametric decomposition of the likelihood ratio, but such a decomposition is feasible for a large class of physics analyses, including effective field theory interpretations~\cite{Ghosh:2025fma}. The method can also be extended to incorporate nuisance parameters, following the semi-parametric treatment adopted by ATLAS~\cite{ATLAS:2025clx}. Given the open questions that remain about robustness in realistic settings, we currently err on the side of caution and recommend the Latent Categories approach instead.

While immediately applicable to offline analyses, these efficient
implementations are essential to make NSBI feasible in trigger
environments, and detailed studies in this direction are left for future work. Both approaches compress each event into a small fixed set of numbers: a single learned category index for LC, and a few density-ratio values for MSS. Since these summaries retain most of the information relevant to the measurement, they could be stored in place of full events, allowing inference to be performed later on a
dramatically reduced dataset. Nuisance parameters can still be modeled by varying the simulated samples. The price is a loss of flexibility, since the stored summaries are tied to a predetermined calibration. For online applications such as data scouting, MSS summaries could still be used for subsequent offline inference even when they are not sufficient statistics, with fast Neyman inversion techniques used to guarantee coverage~\cite{Carzon:2025isu}. Coverage is no more of a concern for the Latent Categories approach than for a traditional histogram-based analysis.

The code to implement our hybrid NSBI techniques and to reproduce our experiments can be found on \href{https://github.com/benevedes/hybrid-nsbi}{GitHub}. The Gaussian dataset can be regenerated with this code. The off-shell Higgs samples were produced with the modified MCFM 10.3 generator of Ref.~\cite{Ghosh:2025fma}, which is publicly available at \href{https://github.com/taehyounpark/MCFM-10.3}{https://github.com/taehyounpark/MCFM-10.3}; the samples themselves are available at \href{https://portal.nersc.gov/cfs/m5295/ML4FP2026/NSBIData/h4l\_data/}{this address}.

\begin{acknowledgments}
We thank Andre Frankenthal for comments on an earlier version of the manuscript. 
This research used resources of the National Energy Research Scientific Computing Center (NERSC), a Department of Energy User Facility using NERSC award ERCAP0038314.

\end{acknowledgments}

\appendix

\section{Latent category occupancy}
\label{sec:AppendixLatentCategory}

\begin{figure}
    \centering
    \includegraphics[width=1.0\linewidth]{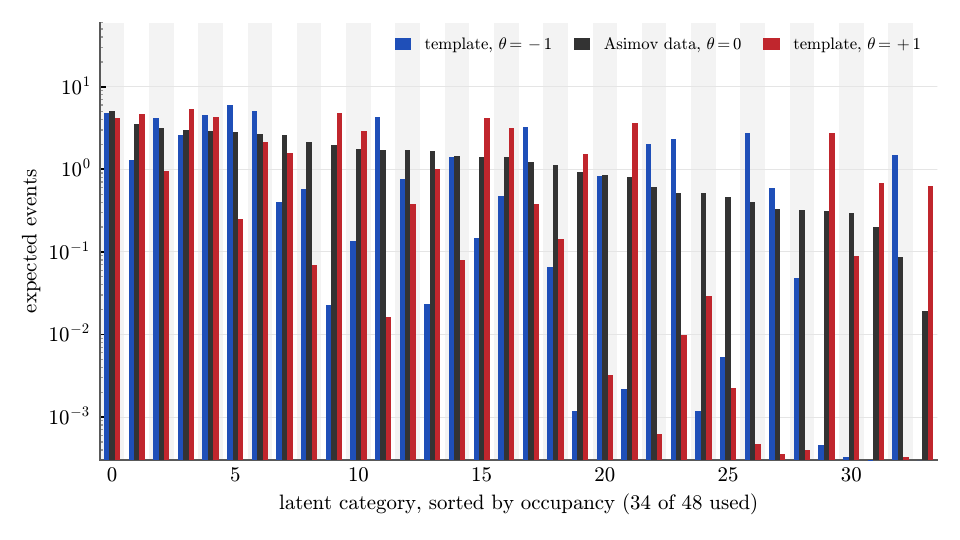}
    \caption{Expected occupancy of each Latent Categories bin for the nominal parameter value $\theta=0$ and the boundary values $\theta=\pm 1$ in the Gaussian case study. The categories, which are intrinsically unordered, are sorted here in order of descending occupancy in the $\theta=0$ case.}
    \label{fig:gaussian_occupancy}
\end{figure}
\begin{figure}
    \centering
    \includegraphics[width=1.0\linewidth]{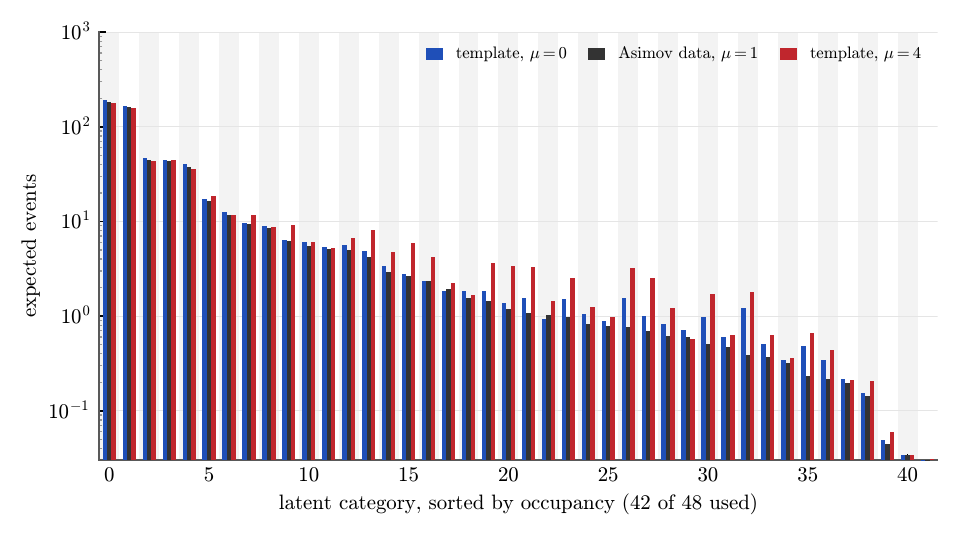}
    \caption{Expected occupancy of each Latent Categories bin for the nominal parameter value $\mu=1$ and the boundary values $\mu=0,4$ in the physics case study. The categories, which are intrinsically unordered, are sorted here in order of descending occupancy in the $\mu=1$ case.}
    \label{fig:physics_occupancy}
\end{figure}

This appendix contains more details on the learned latent categories of the LC approach in each of the two case studies considered in the main text.
In particular, Fig.~\ref{fig:gaussian_occupancy} shows the occupancies of the $34$ bins that the embedding utilizes in the Gaussian case study, and Fig.~\ref{fig:physics_occupancy} shows these occupancies for the $42$ bins used by the embedding in the physics case study.

The categories have no intrinsic ordering, so we sort them in decreasing order of occupancy at the nominal parameter value ($\theta=0$ for the Gaussian case and $\mu = 1$ for the off-shell Higgs case), and red (blue) bars show the occupancy of each of these categories for events generated with the maximum (minimum) value of the parameter. The jaggedness of the bin heights for $\theta = \pm 1$ in the Gaussian case emphasizes this lack of natural ordering, and the sharp distinctions between the occupancies of the three bars for each parameter value show that this binning contains information useful to constrain $\theta$. 

The occupancies are more similar to each other in the off-shell Higgs case study, reflecting the fact that individual events in this case study carry less information about $\mu$ than individual events in the Gaussian case study carry about $\theta$.

\section{Coverage studies}
\label{sec:AppendixCoverage}

This appendix examines the coverage properties of the LC and MSS estimators (as well as verifying the coverage properties of the histogram and NSBI baselines we consider). These results are complementary to the power results in the main body, since comparisons of power are meaningless without holding coverage properties fixed.

\begin{figure}
    \centering
    \includegraphics[width=\linewidth]{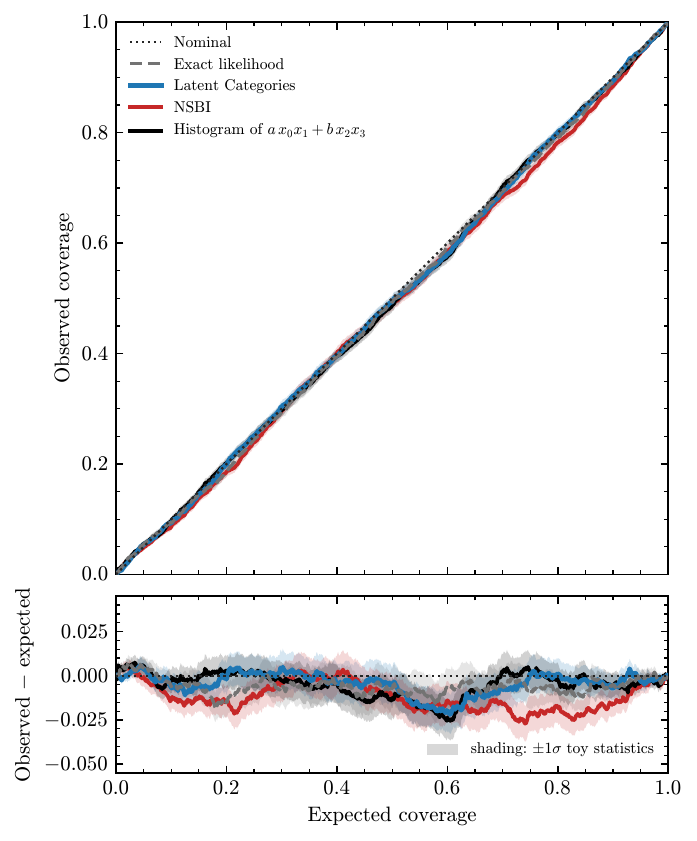}
    \caption{A plot of observed versus expected coverage for the Gaussian case study at $\theta = 0$. The dotted black line shows nominal coverage, the dashed gray line shows the coverage attained using the exact likelihood, and each of the solid lines corresponds to coverage of one of the procedures considered in the main text. The shading corresponds to the $1 \sigma$ uncertainty in the observed coverage corresponding to the finite number of pseudo-experiments.}
    \label{fig:gaussian_coverage}
\end{figure}

In Fig.~\ref{fig:gaussian_coverage}, we plot observed versus expected coverage for the Gaussian case study, again at $\theta = 0$. Coverage is estimated using asymptotic likelihood-ratio test intervals in $2000$ pseudo-experiments, with uncertainties on the observed coverage calculated using the $1 \sigma$ uncertainty on a binomial proportion. We can see that all methods considered are consistent with achieving nominal coverage in this case study.

\begin{figure}
    \centering
    \includegraphics[width=\linewidth]{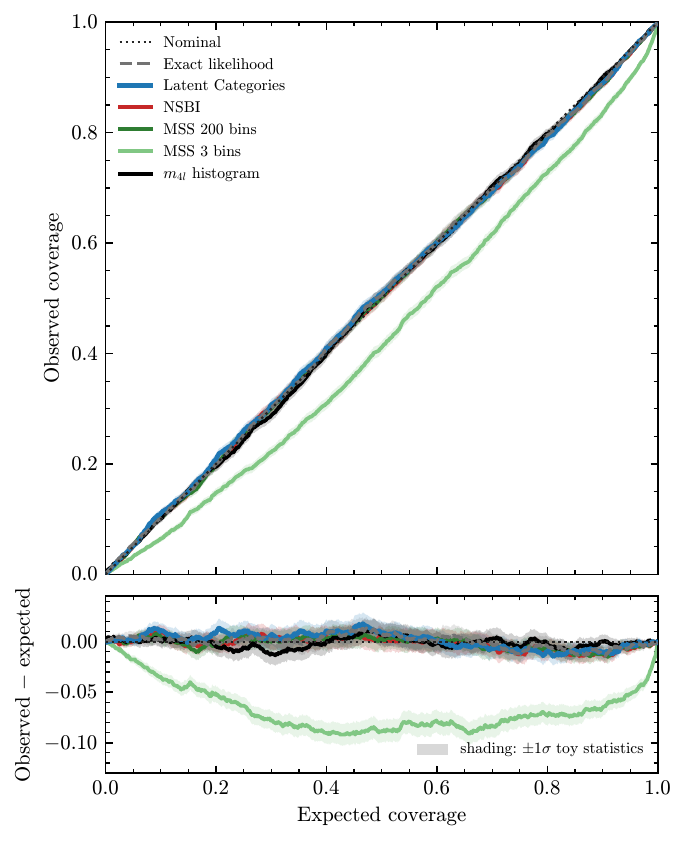}
    \caption{A plot of observed versus expected coverage for the physics case study at $\mu = 10$. The dotted black line shows nominal coverage, the dashed gray line shows the coverage attained using the exact likelihood, and each of the solid lines corresponds to coverage of one of the procedures considered in the main text (except for the ``MSS $3$ bins'' procedure, which is only considered in this appendix). The shading corresponds to the $1 \sigma$ uncertainty in the observed coverage corresponding to the finite number of pseudo-experiments.}
    \label{fig:physics_coverage}
\end{figure}

Then, in Fig.~\ref{fig:physics_coverage}, we plot observed versus expected coverage for the off-shell Higgs case study, this time at $\mu = 10$.\footnote{We choose $\mu = 10$, rather than the nominal $\mu = 1$ considered for the power studies in the main body, to ensure that we compare asymptotic coverage in a regime where the asymptotics are reliable. At $\mu = 1$, we observe small departures of even the exact matrix-element likelihood from nominal coverage due to unreliability of the asymptotics at this parameter value, so we focus on $\mu = 10$ to disentangle the method dependence from unreliability of the asymptotics.} Coverage is again estimated using asymptotic likelihood-ratio test intervals in $2000$ pseudo-experiments, with uncertainties on the observed coverage calculated using the $1 \sigma$ uncertainty on a binomial proportion. 

In particular, we consider MSS with two different binnings: the fine binning used in the main body with $200$ bins and an extremely coarse binning with only $3$ bins. As in the Gaussian case study, all methods used in the main body are consistent with nominal coverage. However, with only $3$ bins, MSS undercovers: this is because with such a coarse binning, the bin index of a given event is not a sufficiently good approximation to a sufficient statistic, and there is a resultant bias. 

This bias is approximately yield-independent, but the uncertainties on $\mu$ decrease with yield, so coverage with this coarse-binned MSS estimator would deteriorate catastrophically at larger yields. We find that the $200$ bin prescription used throughout achieves nominal coverage even at ten times the nominal luminosity, so this implementation of MSS is suitably robust for this case study, but due to the empirical nature of this statement and its potential problem dependence, we recommend LC instead.
\FloatBarrier

\bibliography{references}

\end{document}